\documentclass[%
 reprint,
superscriptaddress,
showpacs,
 amsmath,amssymb,
 aps,
floatfix,
]{revtex4-1}

\usepackage{dcolumn}
\usepackage{bm}

\usepackage[dvipdfmx]{graphicx,color}
\usepackage{color}

\usepackage{ulem} 

\begin{document}


\title{Enhanced charge excitations in electron-doped cuprates by resonant inelastic x-ray scattering}

\author{Takami Tohyama}
\email{tohyama@rs.tus.ac.jp}
\affiliation{Department of Applied Physics, Tokyo University of Science, Tokyo 125-8585, Japan}

\author{Kenji Tsutsui}
\affiliation{Quantum Beam Science Center, Japan Atomic Energy Agency, Hyogo 679-5148, Japan}

\author{Michiyasu Mori}
\affiliation{Advanced Science Research Center, Japan Atomic Energy Agency, Tokai, Ibaraki, 319-1195, Japan}

\author{Shigetoshi Sota}
\affiliation{Computational Materials Science Research Team, RIKEN Advanced Institute for Computational Science (AICS), Kobe, Hyogo 650-0047, Japan}

\author{Seiji Yunoki}
\affiliation{Computational Materials Science Research Team, RIKEN Advanced Institute for Computational Science (AICS), Kobe, Hyogo 650-0047, Japan}
\affiliation{Computational Condensed Matter Physics Laboratory, RIKEN, Wako, Saitama 351-0198, Japan}
\affiliation{Computational Quantum Matter Research Team, RIKEN Center for Emergent Matter Science (CEMS), Wako, Saitama 351-0198, Japan}


\date{\today}
             

\begin{abstract}

Resonant inelastic x-ray scattering (RIXS) tuned for the Cu $L$ edge is a possible tool to detect charge excitations in cuprate superconductors. We theoretically investigate the possibility for observing a collective charge excitation by the RIXS. The RIXS process via the intermediate state inevitably makes the spectral weight of charge excitation stronger in electron doping than in hole doping. Electron-hole asymmetry also appears in the dynamical charge structure factor, showing a new enhanced small-momentum low-energy mode in electron doping. These facts indicate a possibility of detecting the new charge mode by RIXS in electron-doped systems.

\end{abstract}
\maketitle

\section{Introduction}

The cuprate superconductor is classified as a doped-Mott insulator. The spin and charge degrees of freedom in the cuprates, thus, exhibit their characteristic dynamics. Recently, resonant inelastic x-ray scattering (RIXS) experiments tuned for the Cu $L$ edge have provided a lot of new insights about their dynamics.~\cite{Ament2011} In hole-doped cuprates, paramagnon excitations near the magnetic zone boundary are less damped~\cite{Dean2015} unlike the theoretical prediction of the $t$-$t'$-$t''$-$J$ model considered to be a canonical model of cuprates.~\cite{Chen2013} In an electron-doped cuprate Nd$_{2-x}$Ce$_x$CuO$_4$ (NCCO), the energy of magnetic excitations identified by RIXS as well as inelastic neutron scattering increases with carrier concentration~\cite{Ishii2014,Lee2014} in contrast with hole-doped systems. The hardening of the magnetic mode is explained by taking into account a three-site correlated hopping process ignored in the $t$-$J$-type model.~\cite{Jia2014}

The charge ordering in the pseudogap state of hole-doped cuprates is a recent hot topic, to which RIXS has contributed significantly.~\cite{Ghiringhelli2012,Tabis2014} Even for the electron-doped NCCO, the presence of charge ordering has been indicated by resonant x-ray scattering.~\cite{Neto2015} A charge-ordering signal is enhanced below a temperature where antiferromagnetic (AFM) fluctuations are developed. The RIXS experiments for NCCO (Refs. 4 and 5) 
have reported a mode that disperses higher in energy than the magnetic one. Whereas the mode has been interpreted as a signature of broken symmetry based on its temperature dependence~\cite{Lee2014} similar to the charge ordering, it has been identified as a charge excitation~\cite{Ishii2014} because of the agreement with incoherent intraband charge excitation observed by Cu $K$-edge RIXS.~\cite{Ishii2005,Basak2012} If the mode is related to charge ordering, a similar one should be observed in $L$-edge RIXS for hole-doped cuprates, but such a mode has not been reported so far. Therefore, clarifying the difference in charge dynamics between hole- and electron-doped cuprates seen by $L$-edge RIXS is crucial for putting forward our understanding of cuprates. In Cu $K$-edge RIXS, such a difference has been attributed to the difference of the screening process.~\cite{Tsutsui2003} In contrast to $K$-edge RIXS, there is no theoretical confirmation about the difference of $L$-edge RIXS between hole- and electron-doped systems.

In this paper, we theoretically investigate electron-hole asymmetry of charge excitations seen by $L$-edge RIXS, focusing on two possible contributions: One is from the RIXS process where charge fluctuation is affected in the intermediate state with the Cu $2p$ core hole, and the other is the difference of dynamical charge structure factor itself. For the former contribution, we start with a two-band model of the CuO$_2$ plane and calculate the RIXS spectrum by using the exact diagonalization technique. The charge spectral weight in the hole-doped system is suppressed in RIXS through the intermediate state where hole carriers are influenced by the core hole. On the other hand, the weight in the electron-doped system is insensitive to the intermediate state because of no electron carrier on the core-hole site. This behavior is also seen in the single-band Hubbard model. These facts naturally explain visibility of charge dynamics in electron-doped cuprates. For the latter contribution, we perform a large-scale density-matrix renormalization-group (DMRG) calculation of dynamical charge structure factor in the Hubbard model with next-nearest-neighbor hopping. We find a new enhanced small-momentum low-energy charge excitation in the electron-doped system. The energy of the excitation is lower than that of spin excitation. This excitation mode is different from the previously confirmed incoherent intraband excitations whose energy is higher than spin excitation. Combining the two contributions, we predict that the $L$-edge RIXS will detect the new charge collective excitation in the electron-doped cuprates.

This paper is organized as follows. The RIXS spectra of the two-band model and the single-band Hubbard model are shown in Sec.~\ref{Sec2} for both the electron and hole dopings. In Sec.~\ref{Sec3}, we calculate the dynamical charge structure factor of the single-band Hubbard model obtained by dynamical DMRG and discuss a new enhanced small-momentum low-energy charge excitation in the electron-doped system. Finally, a summary is given in Sec.~\ref{Sec4}.

\section{RIXS spectrum}
\label{Sec2}
We firstly consider a two-band model of the CuO$_2$ plane that contains a Cu $3d_{x^2-y^2}$ orbital and a O $2p$ Wannier orbital with $x^2-y^2$ symmetry,~\cite{Zhang1988} which is equivalent to a standard three-band model of cuprates from which a nonbonding band is removed. Since we focus on Cu $L$-edge RIXS associated with the $3d_{x^2-y^2}$ orbital, the use of the two-band model can be justified. The Hamiltonian reads
\begin{eqnarray}\label{dp}
H_{3d}&=&2T_{pd}\sum_{ij\sigma}\tau_{ij} \left( d^\dagger_{i\sigma}\phi_{j\sigma} +\mathrm{H.c.} \right) \nonumber \\
&+&\Delta\sum_{i\sigma}\phi^\dagger_{i\sigma}\phi_{i\sigma} + U_d\sum_i n^d_{i\uparrow}n^d_{i\downarrow},
\end{eqnarray}
where the operator $d_{i\sigma}$ is the annihilation operator for a $3d_{x^2-y^2}$ hole with spin $\sigma$ at site $i$, $n^d_{i\sigma}=d^\dagger_{i\sigma}d_{i\sigma}$, $\phi_{i\sigma}$ represents the annihilation operator of the symmetric O $2p$ orbital, $\tau_{ij}=\frac{1}{N}\sum_\mathbf{k}\beta_\mathbf{k}^{-1}e^{i\mathbf{k}\cdot(\mathbf{R}_i-\mathbf{R}_j)}$ with the position vector $\mathbf{R}_i$ and $\beta_\mathbf{k}=[\sin^2(k_x/2)+\sin^2(k_y/2)]^{-1/2}$, $\Delta$ is the level separation between $3d_{x^2-y^2}$ and $2p$ orbitals, and $U_d$ is the Coulomb repulsion of the Cu $3d_{x^2-y^2}$ orbital. We take $T_{pd}=1$~eV, $\Delta=3$~eV, and $U_d=8$~eV, which are typical values appropriate for cuprates.

In RIXS, tuning the polarization of incident and outgoing photons, we can separate excitation with the change of total spin by one ($\Delta S=1$) and excitation with no change of total spin ($\Delta S=0$)~\cite{Haverkort2010,Igarashi2012,Kourtis2012} (see Appendix~\ref{A}). The two excitations can be defined as
\begin{eqnarray}\label{IDS0}
I^{\Delta S=0}_\mathbf{q} \left(\Delta \omega \right) &=& \sum\limits_f \left| \left\langle f \right|N^j_\mathbf{q} \left| 0 \right\rangle \right|^2 \delta \left( \Delta \omega  - E_f + E_0 \right)\\
\label{IDS1}
I^{\Delta S=1}_\mathbf{q} \left(\Delta \omega \right) &=& \sum\limits_f \left| \left\langle f \right|S^j_\mathbf{q} \left| 0 \right\rangle \right|^2 \delta \left( \Delta \omega  - E_f + E_0 \right)
\end{eqnarray}
with $S^j_\mathbf{q}=(B^j_{\mathbf{q}\uparrow\uparrow}-B^j_{\mathbf{q}\downarrow\downarrow})/2$, $N^j_\mathbf{q}=B^j_{\mathbf{q}\uparrow\uparrow}+B^j_{\mathbf{q}\downarrow\downarrow}$, and
\begin{equation}\label{Bqw}
B^j_{\mathbf{q}\sigma'\sigma}=\sum_l e^{-i\mathbf{q}\cdot\mathbf{R}_l} d^\dagger_{l\sigma'}\frac{1}{\omega_\mathrm{i}-H_l^j+E_0+i\Gamma} d_{l\sigma},
\end{equation}
where $\left|0 \right\rangle$ and $\left|f \right\rangle$ represent the ground state and final state with energy $E_0$ and $E_f$, respectively; $j$ is the total angular momentum of Cu $2p$ with either $j=1/2$ or $j=3/2$; $\mathbf{R}_l$ is the position vector at site $l$; $\omega_\mathrm{i}$ is the incident photon energy; $\Gamma$ is the relaxation time of the intermediate state; and $H_l^j=H_{3d}+U_\mathrm{c} \sum_\sigma n_{l\sigma} + \varepsilon_j$ with $U_\mathrm{c}$ and $\varepsilon_j$ being the Cu $2p$-$3d$ Coulomb interaction and energy level of Cu $2p$, respectively. Here, we assume the presence of a core hole at site $l$. We note that, if we take a fast-collision approximation assuming $\Gamma$ is very large,~\cite{Ament2011} $N^j_\mathbf{q}$ and $S^j_\mathbf{q}$ become the standard charge operator $N_\mathbf{q}$ and the $z$ component of spin operator $S_\mathbf{q}^z$ of a $3d_{x^2-y^2}$ hole, respectively.

\begin{figure}
\includegraphics[width=0.45\textwidth]{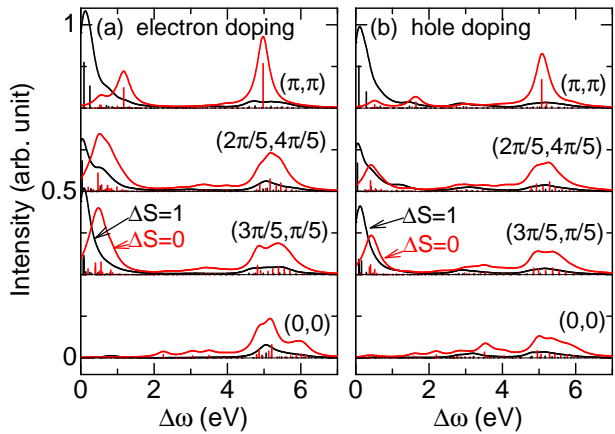}
\caption{(Color online)
RIXS spectra on the ten-unit-cell cluster of the two-band model (\ref{dp}). (a) Electron doping and (b) hole doping with $x=2/10=0.2$.  Black and red colors represent $\Delta S=1$ and $0$ excitations, respectively. The solid curves are obtained by performing a Lorentzian broadening with a width of 0.2~eV on the delta functions denoted by vertical bars. Parameters are $U_d=8$, $\Delta=3$, $T_{pd}=1$, $U_\mathrm{c}=9$, $\Gamma=0.5$ in units of eV. Incident photon energies for RIXS are set to the edges of absorption spectra.
}
\label{fig1}
\end{figure}

In order to calculate Eqs.~(\ref{IDS0}) and (\ref{IDS1}), we use the Lanczos-type exact diagonalization technique on a $\sqrt{10}\times\sqrt{10}$-unit-cell cluster under periodic boundary conditions. We set $U_\mathrm{c}=9$~eV and $\Gamma=0.5$~eV. The incident-photon energy $\varepsilon_\text{i}$ is tuned to the energy of the Cu $2p$ absorption peak that is calculated separately (not shown). Figure~\ref{fig1}(b) shows $I^{\Delta S=0}_\mathbf{q} \left(\Delta \omega \right)$ and $I^{\Delta S=1}_\mathbf{q} \left(\Delta \omega \right)$ for the carrier concentration $x=2/10=0.2$ in hole doping.~\cite{q_range} The low-energy excitation below 1~eV is dominated by the $\Delta S=1$ excitation, while the $\Delta S=0$ excitation is higher in energy. These behaviors are consistent with calculated RIXS of a single-band Hubbard model.~\cite{Jia2014} The results of the $x=0.2$ electron doping are shown in Fig.~\ref{fig1}(a). The distribution of $I^{\Delta S=0}_\mathbf{q}$ and $I^{\Delta S=1}_\mathbf{q}$ is similar to the hole-doped case, but the spectral weight of $I^{\Delta S=0}_\mathbf{q}$ is larger in the electron-doped case. This means that the charge excitation is more visible in electron doping than in hole doping.

\begin{figure}
\includegraphics[width=0.45\textwidth]{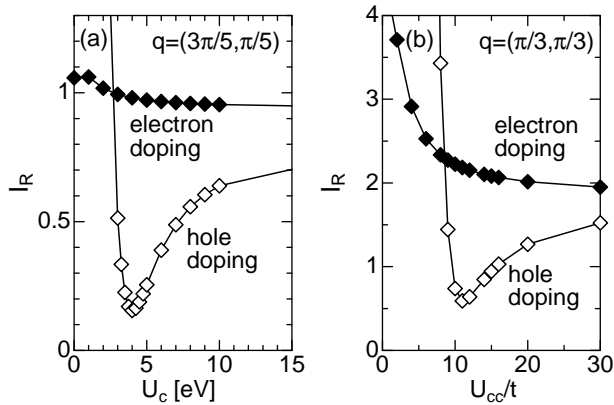}
\caption{
(a) The core-hole potential $U_\mathrm{c}$ dependence of spectral-weight ratio $I_\text{R}=\tilde{I}^{\Delta S=0}_\mathbf{q}/\tilde{I}^{\Delta S=1}_\mathbf{q}$ at $\mathbf{q}=(3\pi/5,\pi/5)$ for the ten-unit-cell cluster of the two-band model, where $\tilde{I}$s are obtained by integrating the weight up to $\Delta\omega=2$~eV in Fig.~\ref{fig1}. Filled and open diamonds represent the $x=0.2$ electron- and hole-doped cases, respectively.  (b) The same as (a) but for the 18-site single-band Hubbard with $U=10t$ and $\Gamma=t$ at $\mathbf{q}=(\pi/3,\pi/3)$. The weights are integrated up to $\Delta\omega=6t$. Filled and open diamonds represent $x=2/18=0.11$ electron-doped and hole-doped cases, respectively.}
\label{fig2}
\end{figure}

The electron-hole asymmetry of $I^{\Delta S=0}_\mathbf{q}$ partly comes from the RIXS process where charge fluctuation is affected in the intermediate state with Cu $2p$ and Cu $3d$ Coulomb interaction. In order to clarify this, we calculate integrated spectral weight $\tilde{I}^{\Delta S=0}_\mathbf{q}$ ($\tilde{I}^{\Delta S=1}_\mathbf{q}$) up to 2~eV for the $\Delta S=0$ ($\Delta S=1$) excitation and take the ratio $I_\text{R}=\tilde{I}^{\Delta S=0}_\mathbf{q}/\tilde{I}^{\Delta S=1}_\mathbf{q}$. We plot the ratio at $\mathbf{q}=(3\pi/5,\pi/5)$ as a function of $U_\mathrm{c}$ in Fig.~\ref{fig2}(a).~\cite{note} In hole doping, $I_\text{R}$ decreases with decreasing $U_\mathrm{c}$ showing a minimum at $U_\mathrm{c}\sim \Delta=3$~eV, while $I_\text{R}$ is less dependent on $U_\mathrm{c}$ in electron doping. Smaller charge spectral weight in hole doping shown in Fig.~\ref{fig1} ($U_\mathrm{c}=9$~eV) is related to the decrease of $I_\text{R}$ as explained below.

We also performed the exact diagonalization calculation of $I^{\Delta S=0,1}_\mathbf{q}(\Delta \omega)$ for the single-band Hubbard model given by
\begin{equation}
H_{tU}=-t\sum_{i\delta\sigma} c^\dagger_{i\sigma} c_{i+\delta\sigma} + U\sum_i n_{i\uparrow}n_{i\downarrow},
\label{singleH}
\end{equation}
where $c^\dagger_{i\sigma}$ is the creation operator of an electron with spin $\sigma$ at site $i$, number operator $n_{i\sigma}=c^\dagger_{i\sigma}c_{i\sigma}$, $i+\delta$ represents the four nearest-neighbor sites around site $i$, and $t$ and $U$ are the nearest-neighbor hopping and on-site Coulomb interaction, respectively. In calculating $L$-edge RIXS, we introduced an attractive  Coulomb interaction $U_\mathrm{cc}$ between the core hole and the valence electron.

Figure~\ref{fig2}(b) shows $U_\mathrm{cc}$ dependence of the ratio $I_R$ defined by the integrated weight up to $\Delta \omega=6t$ at $\mathbf{q}=(\pi/3,\pi/3)$. With decreasing $U_\mathrm{cc}$ down to $U_\mathrm{cc}\sim U$, $I_R$ in hole doping decreases, which is similar to Fig.~\ref{fig2}(a). Since there is no electron-hole asymmetry in the single-band Hubbard model, the asymmetric $I_R$ dependence is purely caused by the effect of $U_\mathrm{cc}$ working at the intermediate state. In electron doping, the incident photon only kicks the site where there is no electron carrier, i.e., the undoped site, since an electron-doped site has no empty state to accommodate a core electron. In this case, the intermediate state of RIXS does not create additional charge modulation, leading to small $U_\mathrm{cc}$ dependence on $I_R$. In hole doping, we have to take into account both undoped and hole-doped sites. When the undoped site is excited by the photon, double occupation of electrons is induced with energy cost $U-2U_\mathrm{cc}$. On the other hand, the hole-doped site gives the energy cost $-U_\mathrm{cc}$. In calculating RIXS, we tune the incident photon to the site with lower energy cost. When $U_\mathrm{cc}-U\gg0$, the undoped site is excited as is the case of electron doping, leading to charge fluctuations similar to the electron doping. With decreasing $U_\mathrm{cc}$, the hole-doped site starts to participate in the intermediate state. This eventually suppresses $I_R$ through destructive contribution to the hole-doped sites. The same type of discussion can be applied to the two-band model shown in Fig.~\ref{fig2}(a). The fact that the $\Delta S=0$ excitation is larger in the electron-doped system than in the hole-doped system is thus caused by the RIXS process related to the core-hole Coulomb interaction. We note that the rapid increase of $I_R$ for $U_\mathrm{cc}<U$ is caused by the decrease of spin excitation. 

\begin{figure}
\includegraphics[width=0.45\textwidth]{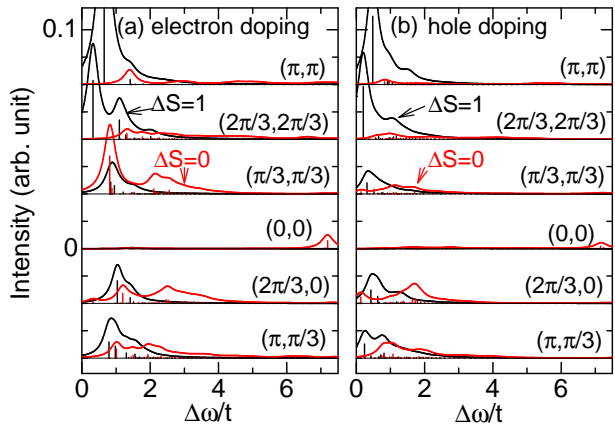}
\caption{(Color online)
RIXS spectra on the 18-site single-band Hubbard cluster with $U=10t$ and $t'=-0.25t$, $U_\mathrm{cc}=15t$, and $\Gamma=t$. (a) Electron doping and (b) hole doping with $x=2/18=0.11$. Black and red colors represent $\Delta S=1$ and $0$ excitations, respectively. The solid curves are obtained by performing a Lorentzian broadening with a width of $0.2t$ on the delta functions denoted by vertical bars. Incident photon energies for RIXS are set to the edges of absorption spectra.}
\label{fig3}
\end{figure}

Next let us consider the effect of the asymmetric electronic state between hole- and electron-doped cuprates on RIXS. Such an asymmetric state is introduced by electron hopping beyond the nearest-neighbor sites. We note that such a long-range hopping is induced by direct oxygen $2p$-$2p$ hopping. Including the next-nearest-neighbor hopping term given by $-t'\sum_{i\delta'\sigma} c^\dagger_{i\sigma} c_{i+\delta'\sigma}$, being $\delta'$ the four next-nearest-neighbor sites, into the single-band Hubbard model (\ref{singleH}), we calculate the RIXS spectra for a $\sqrt{18}\times\sqrt{18}$ periodic lattice with two electrons and two holes ($x=0.11$) as shown in Fig.~\ref{fig3}.~\cite{q_range} Here, we take $t'=-0.25t$. Comparing Fig.~\ref{fig3}(a) and Fig.~\ref{fig3}(b), we find that the $\Delta S=0$ excitation at $\mathbf{q}=(\pi/3,\pi/3)$ is enhanced for electron doping. $I_R$ at $\mathbf{q}=(\pi/3,\pi/3)$ in Fig.~\ref{fig3}(a) is 0.68, which is larger than the case of $t'=0$ with $I_R=0.55$. This increase can be attributed to the asymmetric electronic state. We thus conclude that both the RIXS process and the asymmetric electronic state contribute to enhancing charge excitations in the electron-doped system seen by RIXS, being consistent with the fact that charge excitations have been reported in NCCO (Ref. 4) 
but not yet in hole-doped cuprates though similar charge excitations should exist.

\section{Dynamical charge structure factor}
\label{Sec3}

For making a prediction to RIXS, it is important to clarify characteristics of low-energy charge excitation coming from asymmetric electronic states. We thus calculate dynamical charge structure factor $N(\mathbf{q},\omega)$ for a $6\times 6$ single-band Hubbard cluster with cylindrical geometry, where the $x$ direction is of open boundary condition while the $y$ direction is of periodic boundary condition, by using the dynamical DMRG method. We use $U=8t$, $t'=-0.25t$.~\cite{Jia2014} The numerical method and accuracy are detailed in Appendix~\ref{B}.

\begin{figure}
\includegraphics[width=0.45\textwidth]{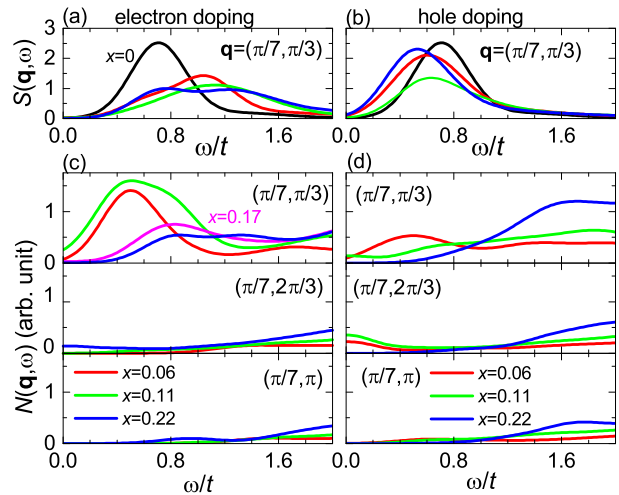}
\caption{(Color online)
Dynamical structure factors in a $6\times 6$ cylindrical Hubbard cluster with $t'=-0.25t$ and $U=8t$. Spin structure factor $S(\mathbf{q},\omega)$ at $\mathbf{q}=(\pi/7,\pi/3)$ for (a) electron doping and (b) hole doping. Charge structure factor $N(\mathbf{q},\omega)$ at various $\mathbf{q}$ for (c) electron doping and (d) hole doping. Black, red, green, purple, and blue lines represent spectra with $x=0$ (half filling), $2/36$, $4/36$, $6/36$, and $8/36$, respectively. A Gaussian broadening width of $0.2t$ is used for the spectral weights.}
\label{fig4}
\end{figure}

In order to confirm doping dependence of spin dynamics,~\cite{Jia2014} we show in Figs.~\ref{fig4}(a) and \ref{fig4}(b) the dynamical spin structure factor $S(\mathbf{q},\omega)$ for electron and hole doping, respectively, at $\mathbf{q}=(\pi/7,\pi/3)$. Whereas the peak position slightly shifts downward with hole doping, the position moves to higher energy with electron doping, as reported by the quantum Monte Carlo calculations.~\cite{Jia2014}

Figure~\ref{fig4}(c) exhibits the doping and momentum dependence of $N(\mathbf{q},\omega)$ for electron doping. At $\mathbf{q}=(\pi/7,\pi/3)$ that is the smallest $\mathbf{q}$ defined in the cluster, the low-energy spectral weights below $\omega=1.6t$ show an interesting doping dependence: The weight increases with increasing $x$ from $x=0.06$ to $x=0.11$ but it decreases with further increase of $x$ followed by an upper shift of spectral weight. More interestingly, the main spectral weight is located in the energy region below spin excitation centered around $\omega=t$. This is a characteristic behavior near half filling in the doped Mott insulator with strong AFM spin correlation.~\cite{Tohyama1996} This also means that this charge mode is different from incoherent intraband charge excitations that have been observed in RIXS experiment for NCCO as a broad and largely dispersive structure above magnetic excitation.~\cite{Ishii2014} In this sense, the charge mode shown in Fig.~\ref{fig4}(c) has not yet been identified in experiment. Since its spectral weights may overlap with those of spin excitations in RIXS experiments, as is expected from our calculations, improving resolution and/or a polarization analysis for outgoing photons is necessary to identify this charge mode in RIXS experiment.

Such a low-energy charge mode is not pronounced in hole doping as shown in Fig.~\ref{fig4}(d). This contrasting behavior may be related to the strength of spin correlation in the spin background. In fact, as is the case of the $t$-$J$-type model,~\cite{Tohyama2004} the AFM spin correlation is large in electron doping than in hole doping near half filling as is seen by the static spin structure factor (see Appendix~\ref{C}). The large AFM spin correlation in the electron-doped system can make a strong spin-charge coupling that induces a collective mode of charge assisted by spin,~\cite{Khaliullin1996} while such an effect is weak in hole doping.

The presence of the low-energy charge mode assisted by spin has been discussed by using $t$-$J$-type models.~\cite{Ishii2014,Khaliullin1996,Tohyama1995,Eder1995} However, its strength depending on carriers has not been discussed by using a realistic parameter set for cuprates. Furthermore, there are only a few works on dynamical charge structure factor of the Hubbard-type model.~\cite{Ishii2005,Grober2000} We thus emphasize that the present results clarify for the first time the low-energy charge mode of electron-doped cuprates by using a realistic Hubbard-type model.

\section{Summary}
\label{Sec4}
In summary, we have theoretically investigated electron-hole asymmetry of charge excitations seen by $L$-edge RIXS and clarified two contributions to the asymmetry. One is related to the RIXS process, in which the charge spectral weight in the hole-doped system is suppressed through the intermediate state where hole carriers are influenced by the core hole, while the weight in the electron-doped system is insensitive to the presence of the intermediate state. Another contribution is due to the asymmetric electronic state inducing a large charge spectral weight in electron doping. These facts naturally explain visible charge excitations in electron-doped NCCO, but not in hole doping. To characterize low-energy charge excitation, we have performed a large-scale DMRG calculation of dynamical charge structure factor in the Hubbard model with the next-nearest-neighbor hopping. We have found an enhanced small-momentum low-energy charge excitation in the electron-doped system. Based on these findings, we predict that the $L$-edge RIXS will detect the new charge collective excitation in the electron-doped cuprates.

\begin{acknowledgments}
We thank G. Khaliullin for fruitful discussions in the initial stage of this work. We also thank K. Ishii, M. Fujita, P. Prelovsek, and O. Sushkov for discussions. This work was supported by the Japan Society for the Promotion of Science, KAKENHI (Grants No. 23340093, No. 24540335, No. 24540387, No. 24360036, No. 26108716, and No. 26287079) and by  HPCI Strategic Programs for Innovative Research (Grants No. hp130007, No. hp140215, and No. hp140078) and Computational Materials Science Initiative from Ministry of Education, Culture, Sports, Science, and Technology.. Numerical calculation was partly carried out at K computer, RIKEN Advanced Institute for Computational Science. T.T., K. T, and M. M. acknowledge the YIPQS program of YITP, Kyoto University.
\end{acknowledgments}

\appendix

\section{RIXS formula}
\label{A}
The spectral weight of Cu $L$-edge RIXS with momentum transfer $\mathbf{q}$ and energy transfer $\Delta \omega$ is given by~\cite{Haverkort2010,Igarashi2012,Kourtis2012}
\begin{equation}\label{RIXS}
I_\mathbf{q}\left(\Delta \omega \right) = \sum_f \left| \left\langle f \right|T_\mathbf{q}\left| 0 \right\rangle \right|^2 \delta \left( \Delta \omega  - E_f + E_0 \right),
\end{equation}
with $T_\mathbf{q}=\sum_j \left( \alpha_N^j N^j_\mathbf{q} + \alpha_S^j S^j_\mathbf{q} \right)$,
where $\left|0 \right\rangle$ and $\left|f \right\rangle$ represent the ground state and final state with energy $E_0$ and $E_f$, respectively, $j$ is the total angular momentum of Cu $2p$ with either $j=1/2$ or $3/2$. The operator $S^j_\mathbf{q}$ induces the change of total spin by one ($\Delta S=1$), but $N^j_\mathbf{q}$ does not charge the total spin ($\Delta S=0$). The operators are given in the main text. The coefficients $\alpha_N^j$ and $\alpha_S^j$ in $T_\mathbf{q}$ are determined by the polarization vector of the incident photon $\mathbf{\epsilon}_\text{i}$ and the outgoing photon $\mathbf{\epsilon}_\text{o}$: $\alpha_N^{j=3/2}=2\alpha_N^{j=1/2}=2(\mathbf{\epsilon}_\text{o}\cdot\mathbf{\epsilon}_\text{i}-\mathbf{\epsilon}^z_\text{o}\mathbf{\epsilon}^z_\text{i})/15$ and $\alpha_S^{j=3/2}=-\alpha_S^{j=1/2}=2i(\mathbf{\epsilon}_\text{o}\times\mathbf{\epsilon}_\text{i})^z/15$.

\section{Dynamical DMRG calculation of dynamical structure factors}
\label{B}
The dynamical charge and spin structure factors, $N(\mathbf{q},\omega)$ and $S(\mathbf{q},\omega)$, are defined as
\begin{eqnarray}
N(\mathbf{q},\omega)&=&-\frac{1}{\pi} \mathrm{Im} \left\langle 0 \right| N_{-\mathbf{q}} \frac{1}{\omega  - H + E_0+i\gamma } N_\mathbf{q} \left| 0 \right\rangle \label{Nqw}\\
S(\mathbf{q},\omega)&=&-\frac{1}{\pi} \mathrm{Im} \left\langle 0 \right| S_{-\mathbf{q}}^z \frac{1}{\omega  - H + E_0+i\gamma } S_\mathbf{q}^z \left| 0 \right\rangle \label{Sqw},
\end{eqnarray}
where $N_\mathbf{q}$ and $S_\mathbf{q}^z$ are the Fourier component of charge operator $n_l=n_{l\uparrow}+n_{l\downarrow}$ and that of the $z$ component of spin operator $S_l^z=(n_{l\uparrow}-n_{l\downarrow})/2$, respectively, with the number operator $n_{l\sigma}$ at site $l$ and spin $\sigma$; $L$ is the system size; $H$ is the Hamiltonian of a given system; and $\gamma$ is a small positive number.

We calculate Eqs.~(\ref{Nqw}) and (\ref{Sqw}) for the Hubbard model with the next-nearest-neighbor hopping $t'$ by dynamical density-matrix renormalization-group (DMRG) method. We use a a $6\times 6$ cluster with cylindrical geometry where the $x$ direction is  of open boundary condition while the $y$ direction is of periodic boundary condition. In the cluster ($L_x=6$ and $L_y=6$), the $y$ component of momentum $\mathbf{q}$ is determined by using a standard translational symmetry, i.e., $q_y=n_y\pi/L_y$ ($n_y=1,2,\cdots,L_y$), but the $x$ component is given by $q_x=n_x\pi/(L+1)$ ($n_x=1,2,\cdots,L_x$) because of the open boundary condition. Defining $l_x$ and $l_y$ as the $x$ and $y$ component of site $l$, respectively, we can write
\begin{eqnarray}
N_\mathbf{q}&=&\sqrt{\frac{2}{(L_x+1)L_y}} \sum_l^L \sin(q_xl_x) e^{-iq_yl_y}n_l\\
S_\mathbf{q}^z&=&\sqrt{\frac{2}{(L_x+1)L_y}} \sum_l^L \sin(q_xl_x) e^{-iq_yl_y}S_l^z.
\end{eqnarray}

In dynamical DMRG, we use three target states for a given $\omega$: For $N(\mathbf{q},\omega)$ ([Eq.~\ref{Nqw}], (i) $\left| 0\right\rangle$, (ii) $N_\mathbf{q} \left| 0 \right\rangle$, and (iii) $(\omega - H + E_0+i\gamma)^{-1} N_\mathbf{q} \left| 0 \right\rangle$. The target state (iii) is evaluated by using a kernel-polynomial expansion method.~\cite{Sota2010} In our kernel-polynomial expansion method, the Lorentzian broadening $\gamma$ is replaced by a Gaussian broadening with half width at half maximum $0.2t$, $t$ being the nearest-neighbor hopping amplitude of the Hubbard model. In our numerical calculations, we divide the energy interval $[0,2t]$ by 100-mesh points and have targeted all of the points at once.

\begin{figure}
\includegraphics[width=0.45\textwidth]{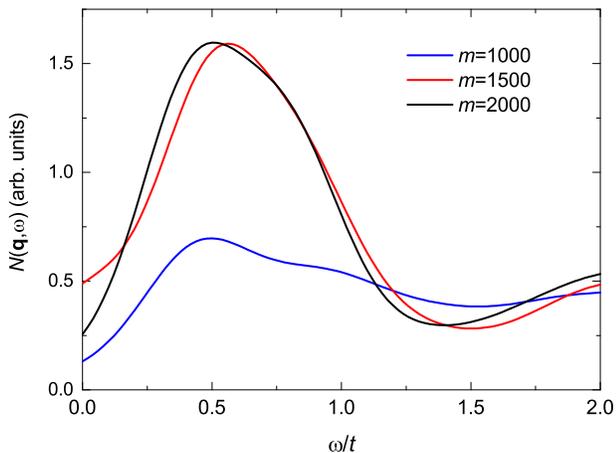}
\caption{(Color online)
The dependence of $N(\mathbf{q},\omega)$ for a give parameter set on the truncation number $m$ in the dynamical DMRG method. A $6\times 6$ cylindrical cluster with $t'=-0.25t$, on-site Coulomb interaction $U=8t$, and $x=4/36$ for electron doping is used. $\mathbf{q}=(\pi/7,\pi/3)$. A Gaussian broadening width of $0.2t$ is used for the spectral weights.}
\label{figA1}
\end{figure}

To perform DMRG, we construct a snakelike one-dimensional chain, and use the truncation number $m=2000$, and resulting truncation error is less than $2\times10^{-3}$. In order to check the accuracy of our results, we examine $m$ dependence of $N(\mathbf{q},\omega)$. Figure~\ref{figA1} shows the results for the carrier concentration $x=4/36$ at $\mathbf{q}=(\pi/7,\pi/3)$. It is clear that the difference between $m=1500$ and $2000$ is small, though the convergence is not perfect. Since it takes 16.5 h by using 3500 nodes in the K computer, RIKEN Advanced Institute for Computational Science, to obtain a single curve of $N(\mathbf{q},\omega)$ with $m=2000$ by our dynamical DMRG code, it is practically difficult to increase $m$ further to perform systematic calculations of the doping and momentum dependence of dynamical structure factors.

\section{DMRG calculation of static spin structure factor}
\label{C}
In the $t$-$t'$-$t''$-$J$ model, it is known that antiferromagnetic spin correlation remains strong in the electron-doped side of the phase diagram.~\cite{Tohyama2004} This is caused by the same sublattice hopping $t'$ affecting the spin background through the stability of the N\'eel-type configuration by the motion of electron carriers. The antiferromagnetically ordered spins can couple to the charge degree of freedom, leading to the low-energy collective charge excitation in the $t$-$J$-type model.~\cite{Khaliullin1996} In order to check whether our $t$-$t'$-$U$ Hubbard model examined by dynamical DMRG has a similar effect on the spin background, we investigate the static spin structure factor defined by
\begin{equation}
S(\mathbf{q})=\left\langle 0 \right| S_{-\mathbf{q}}^z S_\mathbf{q}^z \left| 0 \right\rangle,
\end{equation}
which can measure the strength of antiferromagnetic spin correlation.

\begin{figure*}[htbp]
  \begin{center}
    \begin{tabular}{r}
      \begin{minipage}{0.5\hsize}
        \begin{center}
          \includegraphics[clip, width=13pc]{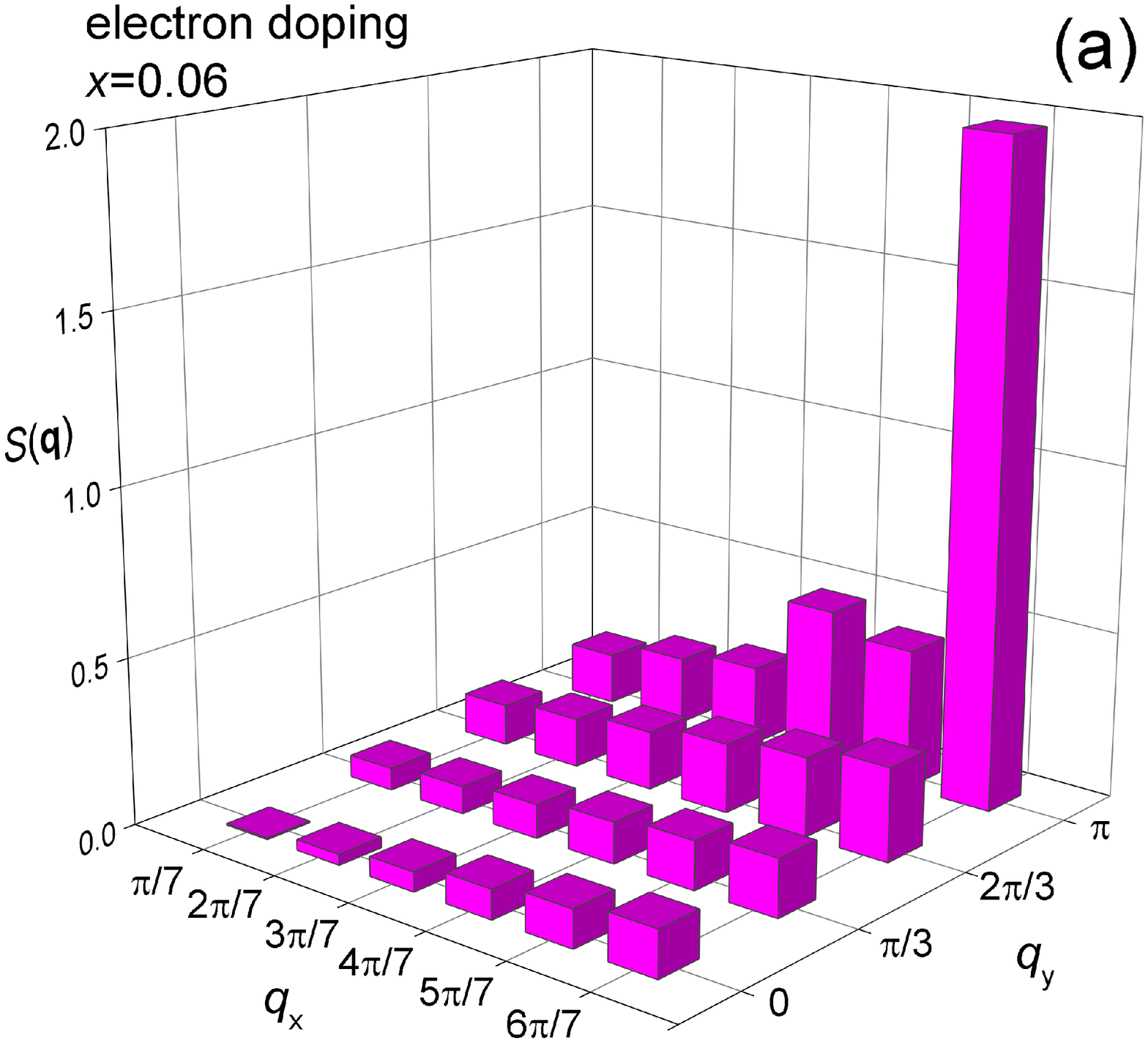}
          \vspace{0.5cm}
        \end{center}
      \end{minipage}
      \begin{minipage}{0.5\hsize}
        \begin{center}
          \includegraphics[clip, width=13pc]{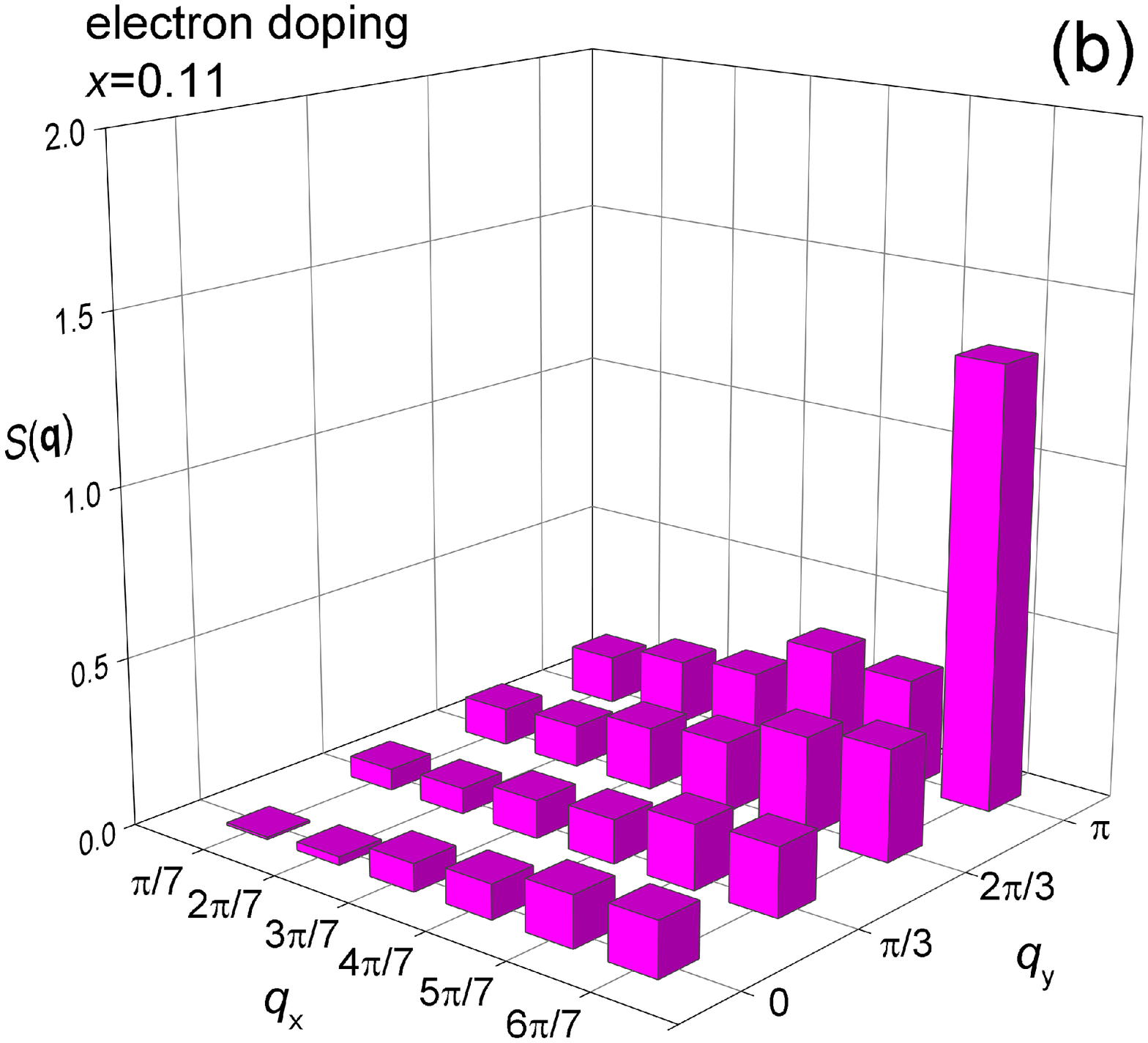}
          \vspace{0.5cm}
        \end{center}
      \end{minipage}
      \\
      \begin{minipage}{0.5\hsize}
        \begin{center}
          \includegraphics[clip, width=13pc]{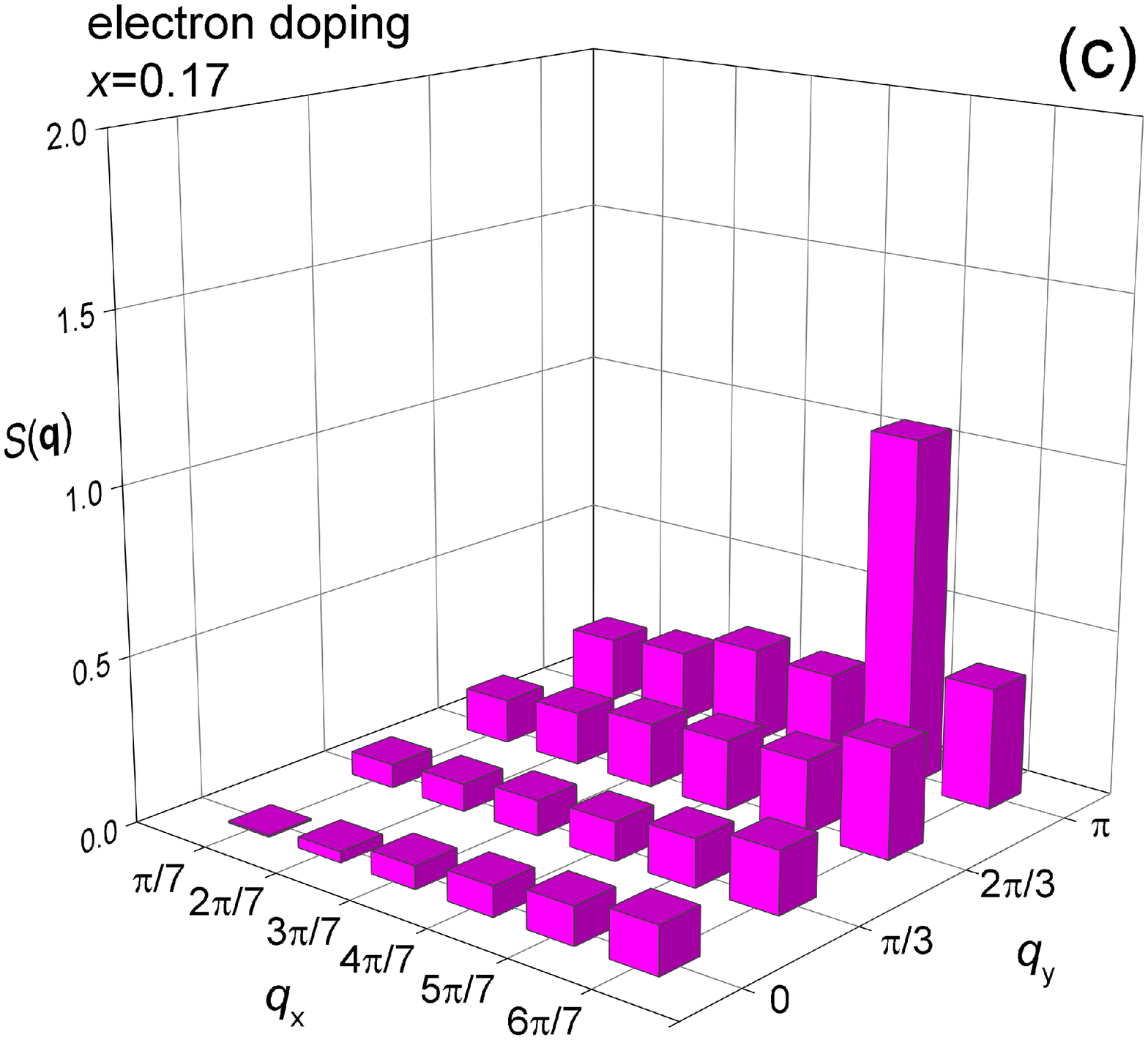}
          \vspace{0.5cm}
        \end{center}
      \end{minipage}      
      \begin{minipage}{0.5\hsize}
        \begin{center}
          \includegraphics[clip, width=13pc]{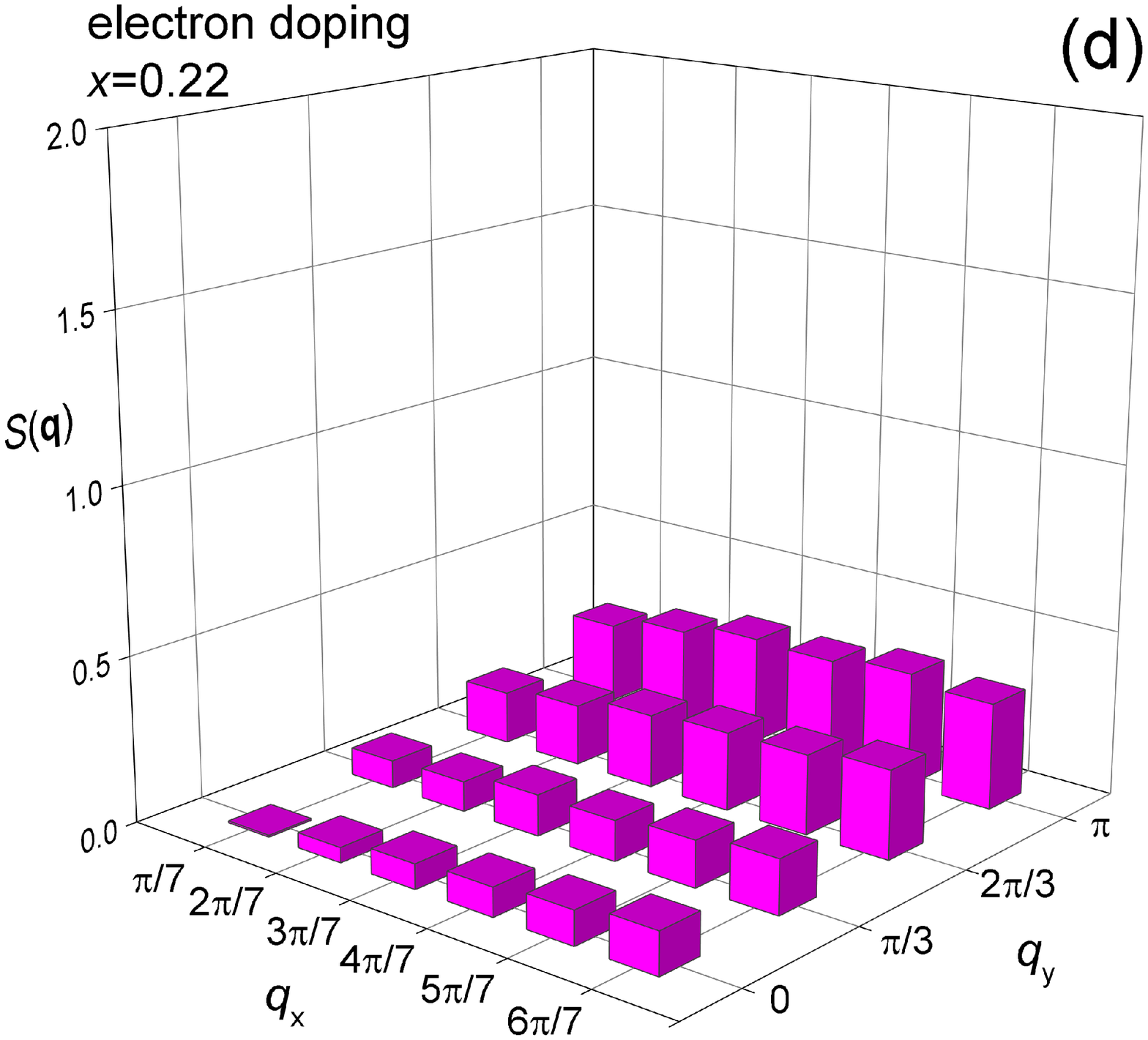}
          \vspace{0.5cm}
        \end{center}
      \end{minipage}      
    \end{tabular}
\caption{(Color online)
The static spin structure factor $S(\mathbf{q})$ for the electron-doped $6\times 6$ cylindrical Hubbard cluster with $t'=-0.25t$ and on-site Coulomb interaction $U=8t$. (a) $x$=0.06, (b) $x$=0.11, (c) $x$=0.17, and (d) $x$=0.22. The height of the bars represents the magnitude of $S(\mathbf{q})$ at a given momentum.}
\label{figA2}
  \end{center}
\end{figure*}

\begin{figure*}[htbp]
  \begin{center}
    \begin{tabular}{r}
      \begin{minipage}{0.5\hsize}
        \begin{center}
          \includegraphics[clip, width=13pc]{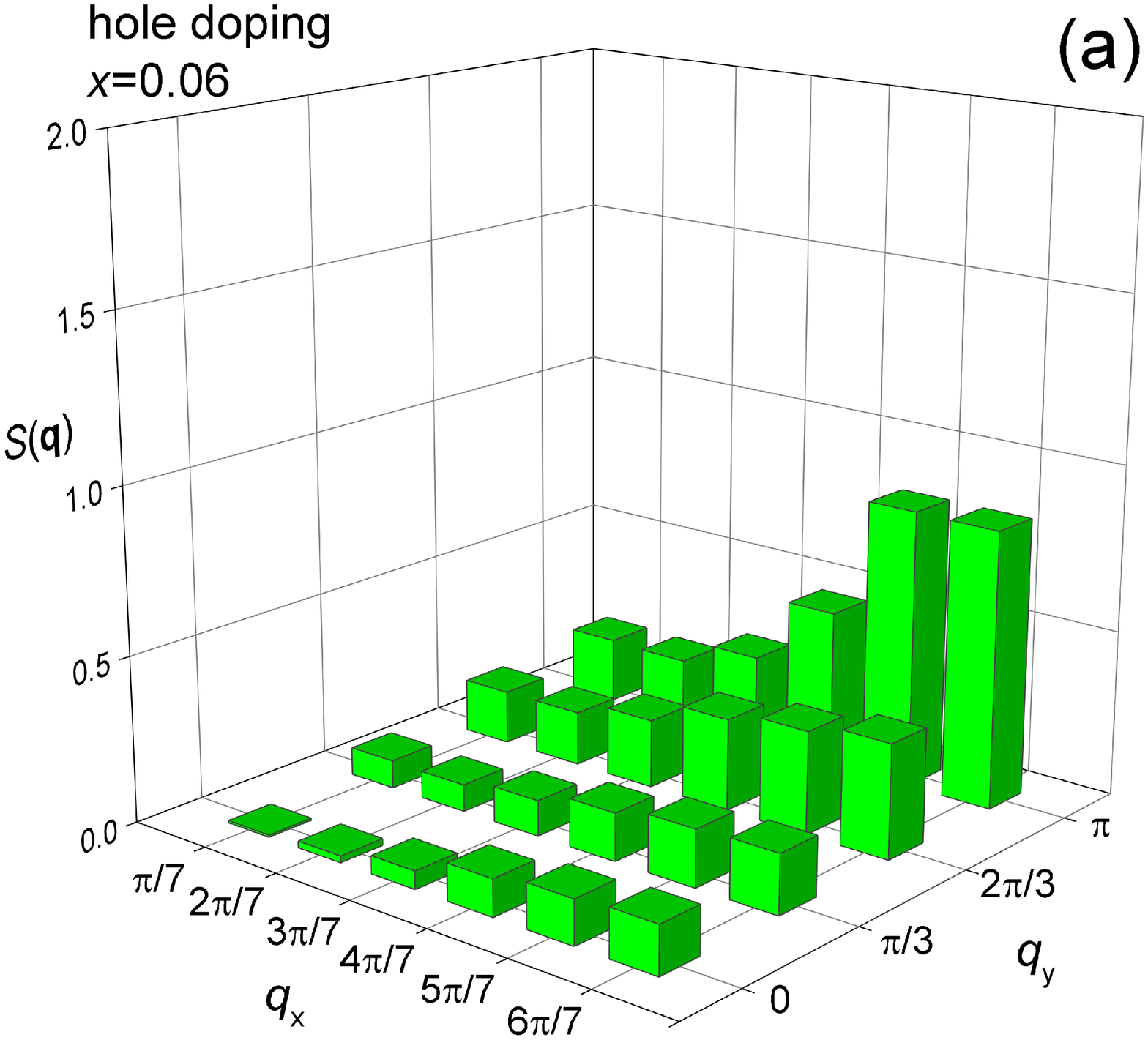}
          \vspace{0.5cm}
        \end{center}
      \end{minipage}
      \begin{minipage}{0.5\hsize}
        \begin{center}
          \includegraphics[clip, width=13pc]{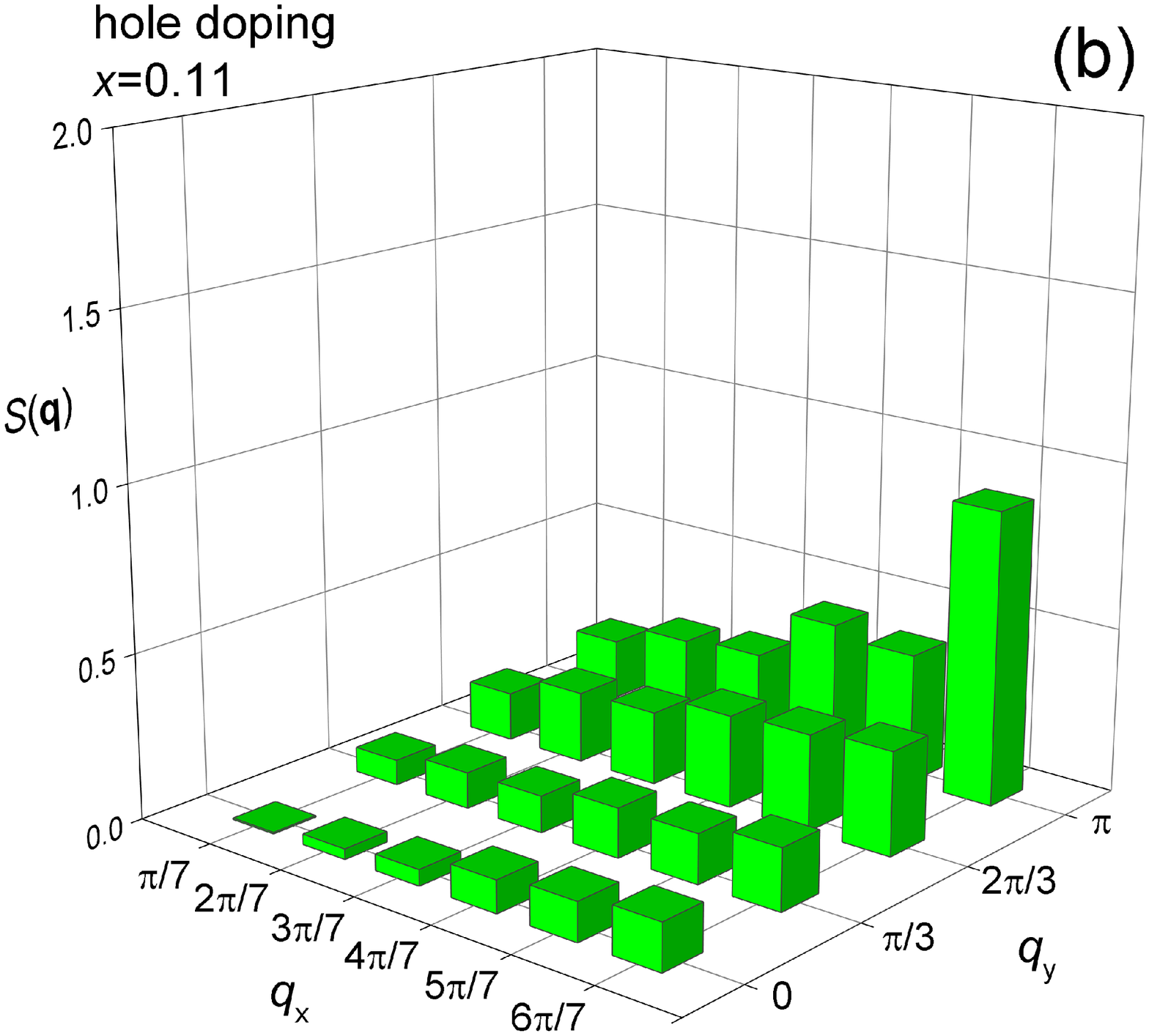}
          \vspace{0.5cm}
        \end{center}
      \end{minipage}
      \\
      \begin{minipage}{0.5\hsize}
        \begin{center}
          \includegraphics[clip, width=13pc]{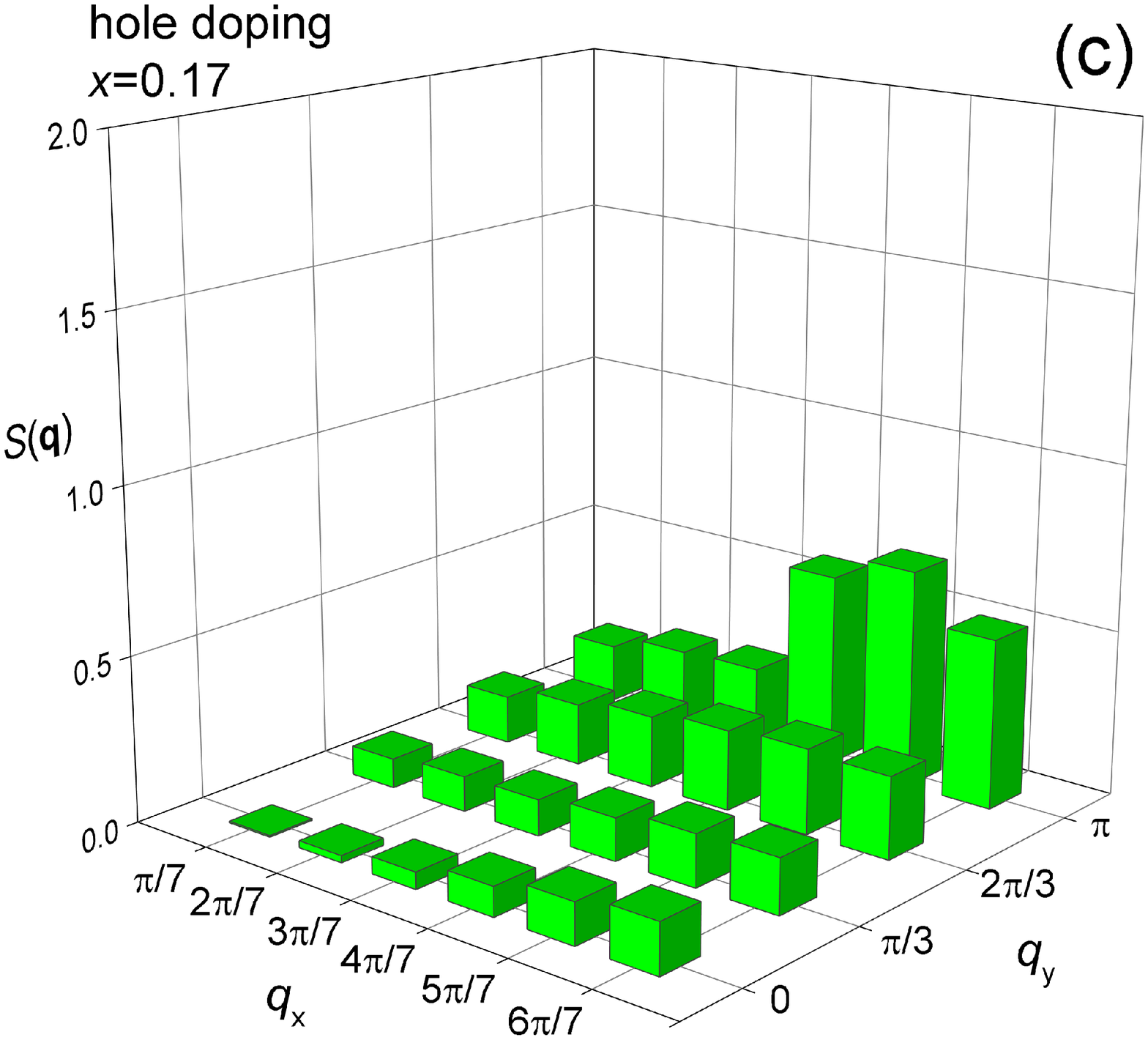}
          \vspace{0.5cm}
        \end{center}
      \end{minipage}      
      \begin{minipage}{0.5\hsize}
        \begin{center}
          \includegraphics[clip, width=13pc]{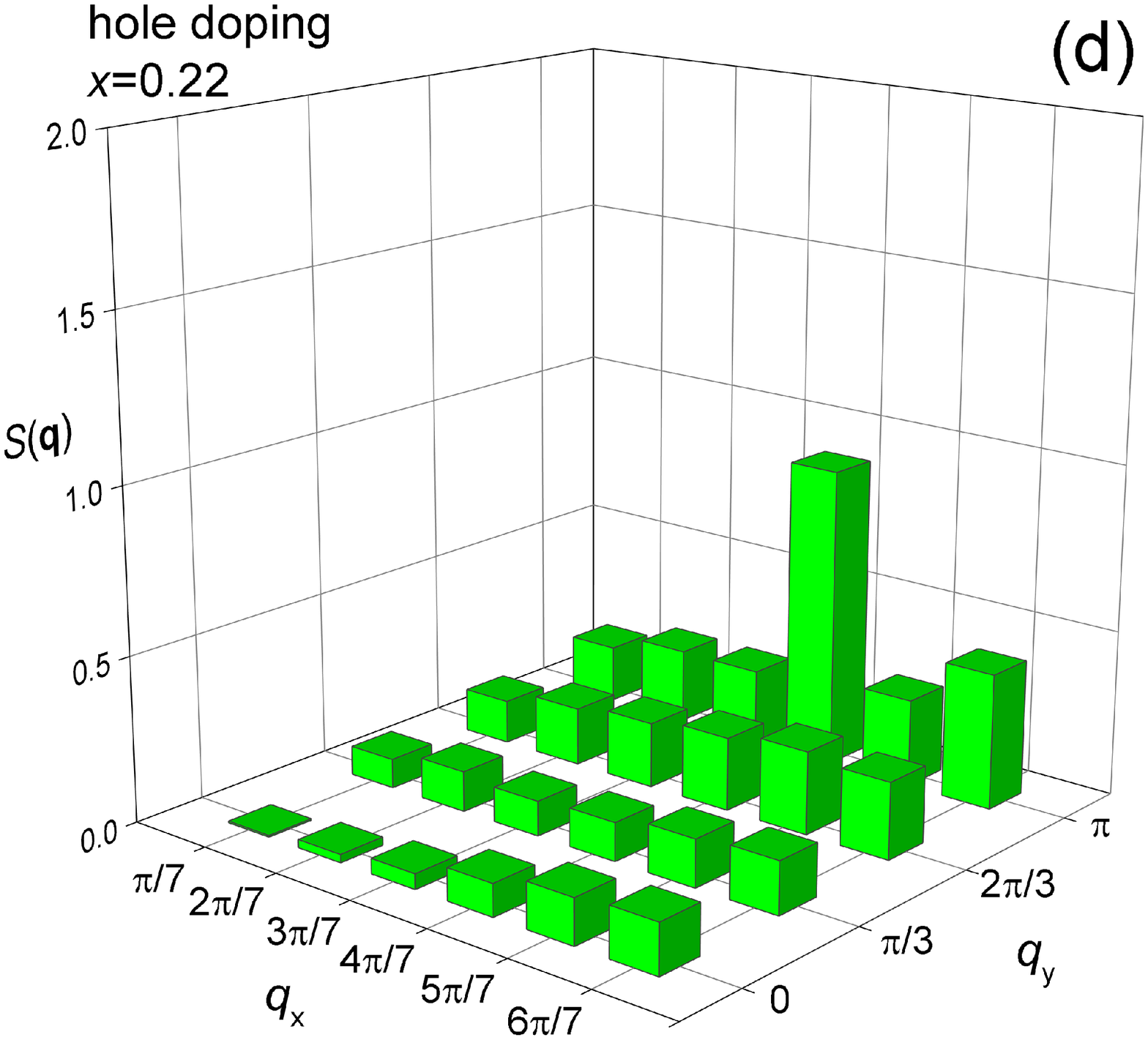}
          \vspace{0.5cm}
        \end{center}
      \end{minipage}      
    \end{tabular}
\caption{(Color online)
The static spin structure factor $S(\mathbf{q})$ for the hole-doped $6\times 6$ cylindrical Hubbard cluster with $t'=-0.25t$ and on-site Coulomb interaction $U=8t$. (a) $x$=0.06, (b) $x$=0.11, (c) $x$=0.17, and (d) $x$=0.22. The height of the bars represents the magnitude of $S(\mathbf{q})$ at a given momentum.}
\label{figA3}
  \end{center}
\end{figure*}

Figures~\ref{figA2} and \ref{figA3} show $S(\mathbf{q})$ for the electron- and hole-doped systems, respectively. It is clear that near half-filling for the electron-doped side [Figs.~\ref{figA2}(a) and \ref{figA2}(b)] the antiferromagnetic spin correlation is dominating in the spin background as evidenced from the large value of $S(\mathbf{q})$ near $\mathbf{q}=(\pi,\pi)$. This is in contrast with the hole doped case in Fig.~\ref{figA3}. We consider that the contrasting behavior of spin correlation contributes to the different behavior of low-energy charge dynamics discussed in the main text.

Another contrasting behavior appears in the over-doped region at $x$=0.22, where an incommensurate spin correlation developed in the hole-doped case (Fig.~\ref{figA3}(d)) but not in the electron-doped case (Fig.~\ref{figA2}(d)). The difference is probably related to the Fermi-surface topology that favors a nesting in the hole-doped side.

\nocite{*}


\begin{thebibliography}{99}
\bibitem{q_range}Large $\mathbf{q}$ spectra to which Cu $L$-edge RIXS cannot have access are shown for complete understanding of RIXS.
\bibitem{note}Within the first-collision approximation, we obtain $I_\mathrm{R}\sim 0.78$ and $\sim0.03$ in electron and hole doping, respectively. We note that these values are very different from those obtained by the present non-perturbative treatment for the core-hole Coulomb interaction.
\end{thebibliography}

\end{document}